# An Investigation into the Correlation between a President's Approval Rating and the Performance of His Party in the Midterm Elections
## By George Pandya

**ABSTRACT**

Over the years, American politics have become increasingly polarized. In today's political landscape, a president cannot easily collaborate with the opposite party and pass legislature. Ideologies between parties have drifted apart to the point that one party generally stonewalls any legislature proposed by the other party. Because of this political landscape, it is paramount for a president to have a majority of his party in Congress. Political parties invest a great deal of time and effort into making sure that first their Presidential candidate wins and is popular, and then their congressional candidates win seats in Congress. In this study, the effect of the former on the latter was investigated – how the president's approval rating influences the number of seats won or lost in Congress during the midterm elections.

The data used was collected from Gallup's nationwide random polling. An analysis of the data yielded the statistically significant linear model $y = -107.423 + 1.594x$, where x is the United States President's approval rating and y is the number of Congressional seats won or lost by the President's party during Midterm Elections. Further analysis yielded a 20% more statistically useful model for approval ratings greater than 50%: $y = -275.461 + 4.37551x$, again where x is the United States President's approval rating and y is the number of Congressional seats won or lost by the President's party during Midterm Elections. An attempt to duplicate the success of the greater than 50% model on a less than 50% model was unsuccessful, as a t test for model utility showed no evidence for a useful model.

As of May 16, 2014, President Barrack Obama's approval rating is 44% according to Gallup's nationwide random polling. Using the originally derived linear model, y = -107.423+1.594x, it can be said with 95% confidence that the Democratic Party, President Obama's party, will lose between 27 and 48 seats in Congress, rounded to the nearest whole seat. During the 2014 Midterm Elections, the Democrats lost 28 seats, as predicted by the model.

These findings point out a striking link between the President's Approval Rating and the performance of his party in Midterm Elections. This model could be used by political advisors to predict the potential outcomes of midterm elections, and advise the tactical doctrine of presidential administrations and political parties going into Midterm Election years,

**STATEMENT OF PROBLEM**

In order for the President to pass legislature more easily, he needs majority support from his party in Congress. In order for the President to have support in Congress, his party needs to win seats in Congress. In order for the President's party to win seats in Congress, it needs to be popular. Because the president is the leader of his party, his approval rating plays a major role in the perceived popularity of this party.

Political advisors strive to find appropriate metrics for predicting various outcomes during election season, allowing them to respond to changes in the political landscape before they happen. The goal of the research was to find a model relating the President's Approval rating and his party's performance during Midterm Elections and use this model to predict the outcome of the 2014 Midterm Elections, as the research and models were established prior to the 2014 Midterm Elections.

## VARIABLES

The independent variable used in this study was the approval rating of the President on the eve of the midterm election, represented by x. The response or dependent variable was the number of seats won or lost by the President's party in Congress during midterm elections, represented by y.

## COLLECTION OF DATA

The data was obtained from Gallup.com. This Gallup data included the following data for every president since 1946: his party, his approval on the eve of the Midterm Elections, and the net gain/loss of his party during that year's election. The approval rating was based on Gallup nationwide random polling, and the seats won/lost were based on Congressional data.

Gallup describes its data collection methods as:

> *Results are based on telephone interviews conducted as part of Gallup Daily tracking July 26-Aug. 1, 2010, with a random sample of 3,544 adults, aged 18 and older, living in all 50 U.S. states and the District of Columbia, selected using random-digit-dial sampling.*
>
> *For results based on the total sample of national adults, one can say with 95% confidence that the maximum margin of sampling error is ±2 percentage points.*
>
> *Interviews are conducted with respondents on landline telephones and cellular phones, with interviews conducted in Spanish for respondents who are primarily Spanish-speaking. Each daily sample includes a minimum quota of 150 cell phone respondents and 850 landline respondents, with additional minimum quotas among landline respondents for gender within region. Landline respondents are chosen at random within each household on the basis of which member had the most recent birthday.*
>
> *Samples are weighted by gender, age, race, Hispanic ethnicity, education, region, adults in the household, cell phone-only status, cell phone-mostly status, and phone lines. Demographic weighting targets are based on the March 2009 Current Population Survey figures for the aged 18 and older non-institutionalized population living in U.S. telephone households. All reported margins of sampling error include the computed design effects for weighting and sample design.*

*In addition to sampling error, question wording and practical difficulties in conducting surveys can introduce error or bias into the findings of public opinion polls.[1]*

**ANALYSIS OF DATA**

The data were first tabulated.

| Year | President | Party | President's Approval Rating Prior to Midterm (Percentage) | Seat Gain/Loss for President's party[2] |
|---|---|---|---|---|
| 1998 | Clinton | Democrat | 66 | 5 |
| 1986 | Reagan | Republican | 63 | -5 |
| 2002 | G. W. Bush | Republican | 63 | 6 |
| 1954 | Eisenhower | Republican | 61 | -4 |
| 1962 | Kennedy | Democrat | 61 | -4 |
| 1970 | Nixon | Republican | 58 | -12 |
| 1990 | G. H. W. Bush | Republican | 58 | -8 |
| 1958 | Eisenhower | Republican | 57 | -47 |
| 1974 | Ford | Republican | 54 | -43 |
| 1978 | Carter | Democrat | 49 | -11 |
| 1994 | Clinton | Democrat | 46 | -53 |
| 2010 | Obama | Democrat | 45 | -63 |
| 1966 | Johnson | Democrat | 44 | -47 |
| 1982 | Reagan | Republican | 42 | -28 |
| 1950 | Truman | Democrat | 39 | -29 |
| 2006 | G. W. Bush | Republican | 38 | -30 |
| 1946 | Truman | Democrat | 33 | -55 |
| | | **Mean** | 51.58823529 | -25.17647059 |
| | | **St. Deviation** | 10.25340689 | 22.73003325 |

And then the data were graphed:

---

[1] "Gallup Daily: Obama Job Approval." *Gallup Daily: Obama Job Approval*. Gallup, 16 May 2014. Web. 17 May 2014. <http://www.gallup.com/poll/113980/Gallup-Daily-Obama-Job-Approval.aspx>.

[2] A negative value corresponds to a net loss of congressional seats by the President's party in Congress

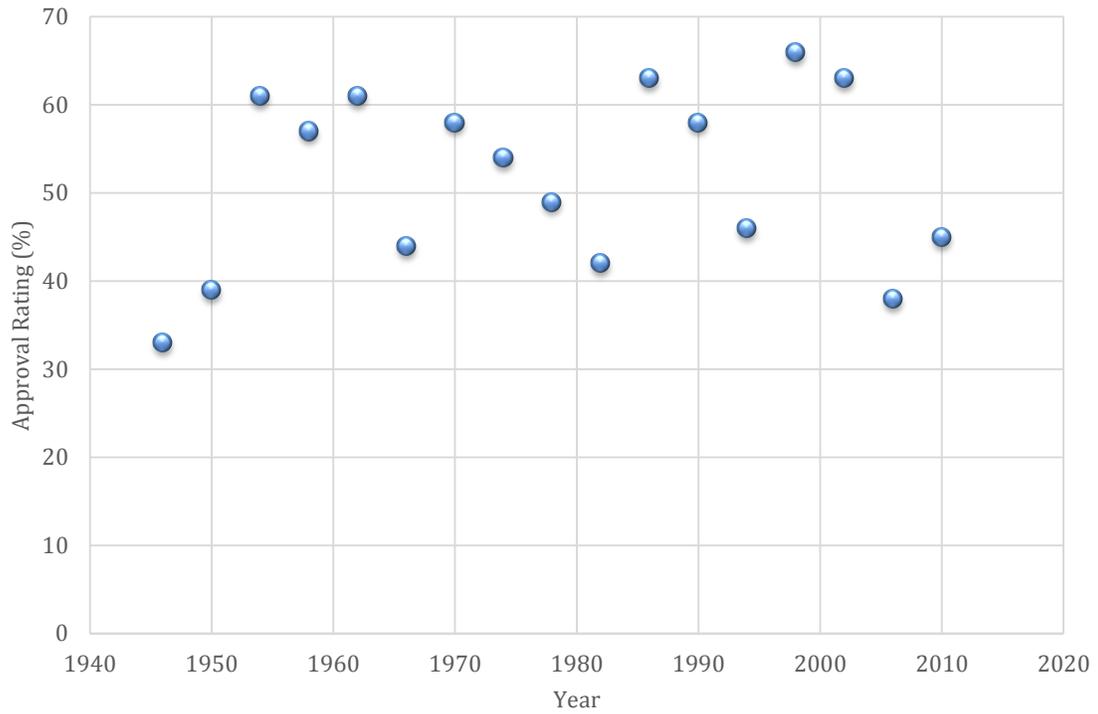

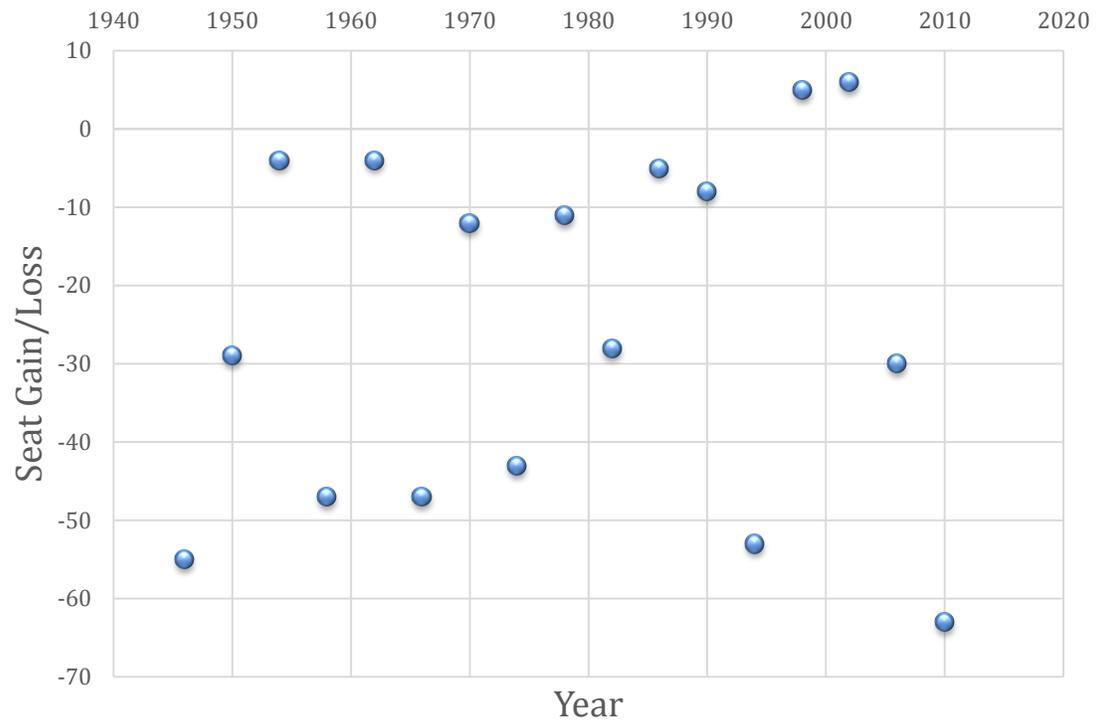

From these data, a few things are immediately clear.

1) The President's political party almost always loses seats in Congress during midterms. There were only two times when the President's party gained congressional seats during midterms and both times, the president had an approval rating greater than 63%.

2) The President's approval rating is centered at 51% with a standard deviation of about 10, which puts most of the approval ratings between 40% and 60%. The seat gain is centered at -25 seats, meaning that on average the President's party loses seats during Midterms. The standard deviation is 22, which means a net loss of seats for the President's party is to be expected over 65% of the time.

A linear regression analysis yielded these data:

| Linear Regression: y=a+bx | | | | | |
|---|---|---|---|---|---|
| a | b*x | $r^2$ | Degrees of Freedom | t | p |
| -107.423 | 1.59428*x | 0.517208 | 15 | 4.00865 | 0.001139 |

The regression line for these data is y = -107.423+1.594x, with a correlation coefficient of 0.517, a moderate correlation.

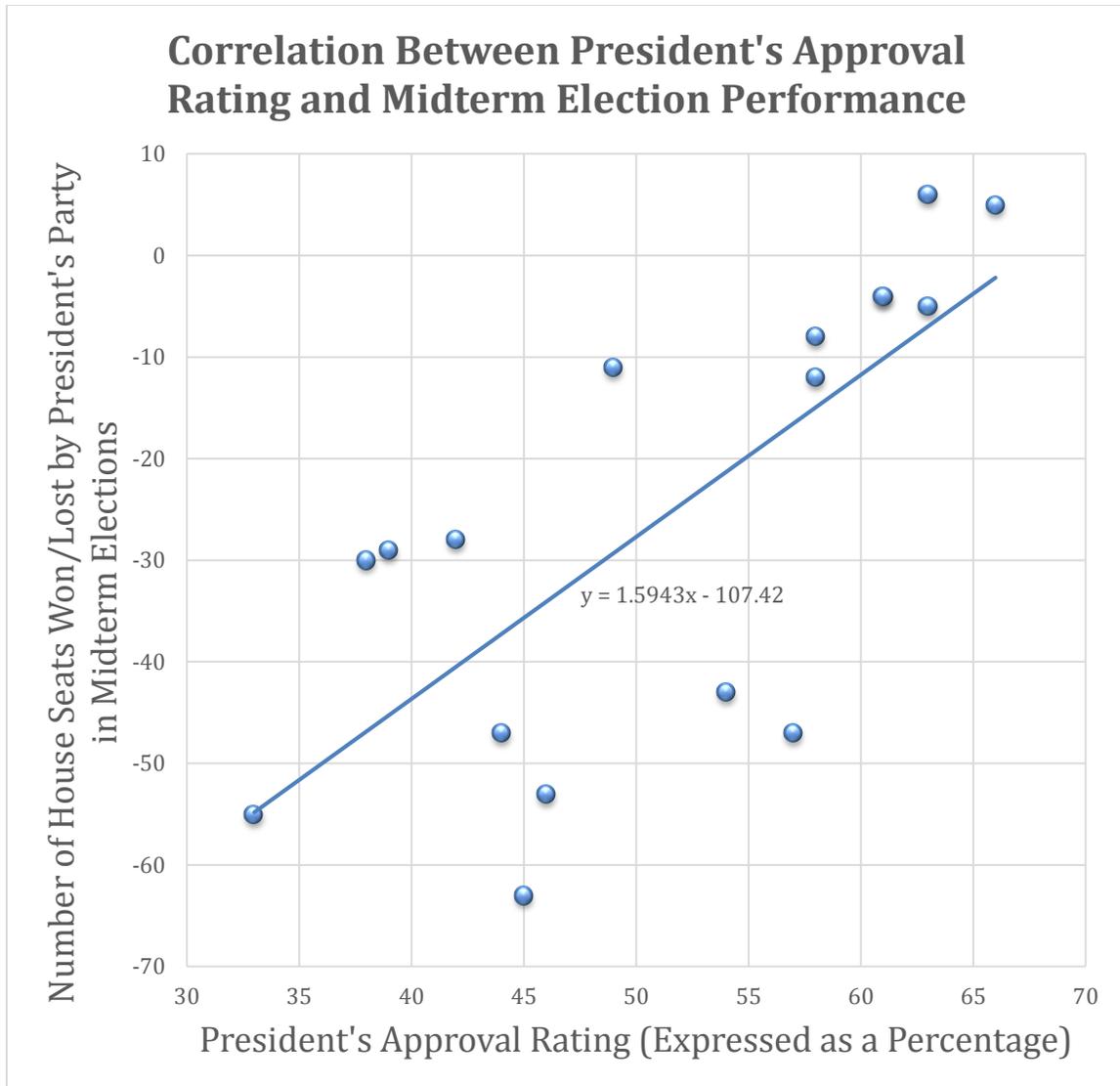

Of course, this correlation only applies if the data is an appropriate fit for the line. A t test for the utility of the linear model with the following hypotheses needs to be performed:

**H₀: The regression line is an improper fit (β=0)**
**Hₐ: H₀ is false**

The above linear regression analysis gives a t critical value of 4.00865 and a corresponding p value of 0.001139. Because the p value, which represents the probability of the null hypothesis being true, is so low, the null hypothesis, H₀, can be rejected with nearly 99.9% confidence.

The line y = -107.423+1.594x is, in fact, a useful model for predicting the seats won or lost by the President's party in Congress based on the approval rating of the United States President.

The second objective of this investigation is to use the above model to predict the outcome of the 2014 Midterm Elections for the Democratic Party based on President Obama's approval rating.

As of October 30, 2014, President Obama's approval rating is 44% according to Gallup's nationwide random polling. Based on these data and the above model, it can be said with 95% confidence that the number of seats lost by the Democratic Party in 2014 will be between 47.88 and 26.67, with a margin of error of about 10.61 seats to 2 significant figures.

To two significant figures, this confidence interval is represented by -37.27±10.61.

**ADDITIONAL ANALYSIS**

In the interest of finding an even more accurate predictor model for the President's party in the Midterm Elections, two more models will be considered – one model for approval ratings greater than 50% and one for approval ratings less than 50%.

**Starting with a model for approval ratings greater than 50%**

| Year | President | Party | President's Approval Rating Prior to Midterm (Percentage) | Seat Gain/Loss for President's party |
|---|---|---|---|---|
| 1998 | Clinton | Democrat | 66 | 5 |
| 1986 | Reagan | Republican | 63 | -5 |
| 2002 | G. W. Bush | Republican | 63 | 6 |
| 1954 | Eisenhower | Republican | 61 | -4 |
| 1962 | Kennedy | Democrat | 61 | -4 |
| **1970** | Nixon | Republican | 58 | -12 |
| **1990** | G. H. W. Bush | Republican | 58 | -8 |
| **1958** | Eisenhower | Republican | 57 | -47 |
| **1974** | Ford | Republican | 54 | -43 |
| | | **Mean** | 60.11111111 | -12.44444444 |
| | | **St. Deviation** | 3.689323937 | 19.33333333 |

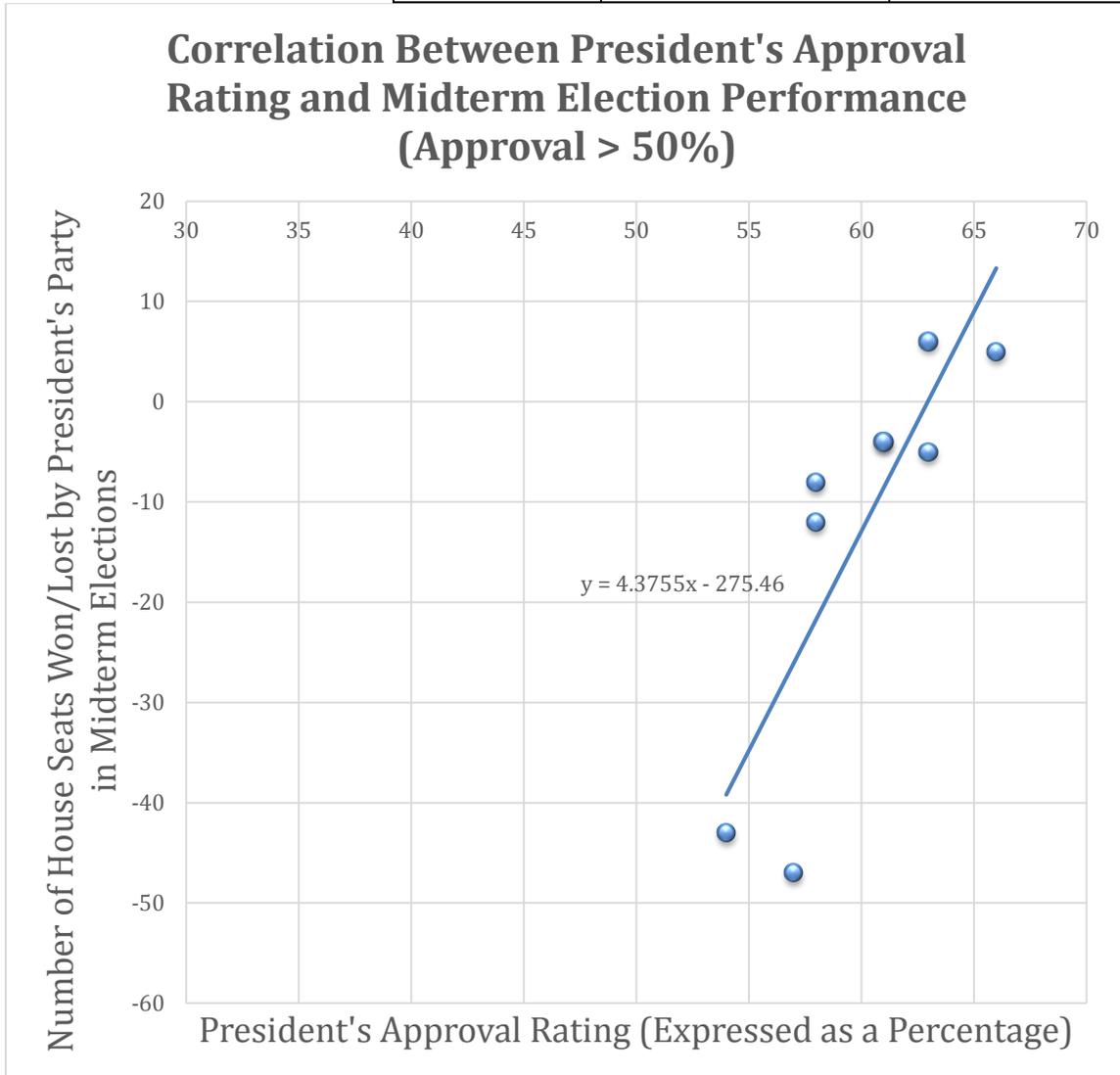

From these data, a few things are immediately clear.

1) The only times when the President's party gained seats in Congress, the president had an approval rating over 63%.

2) The President's approval rating is centered at about 60.1% with a standard deviation of about 3.68. Intriguingly, this means that the approval ratings are predominantly in the upper 50s to lower 60s range. The seat gain is centered at about -12.4 seats, meaning that even when more than half of the public approves of the job being done by the President, on average the President's party loses seats during Midterms, albeit fewer than for Presidents whose approval ratings stand below 50%.

**The next step is to find a model for the data for approval ratings greater than 50%**

| Linear Regression for Approval Ratings Greater than 50% | | | | | |
|---|---|---|---|---|---|
| a | bx | $r^2$ | df | t | p |
| -275.461 | 4.37551*x | 0.697168 | 7 | 4.01437 | 0.005096 |

The regression line for these data is y = -275.461+4.37551x, with a correlation coefficient of about 0.697, showing a moderate to strong correlation. This correlation coefficient amounts to about a 19% stronger correlation than the data set that included all approval ratings. Once again, this correlation only applies if the data is an appropriate fit for the line. Now perform t test for the utility of the linear model with the following hypotheses:

**H₀: The regression line is an improper fit (β=0)**
**Hₐ: H₀ is false**
The above linear regression analysis gives a t critical value of 4.01437 and a corresponding p value of 0.005096. Because the p value is so low, the null hypothesis, H₀, can be rejected with nearly 99.5% confidence.

The line y = -275.461+4.37551x is, in fact, a useful model for predicting the seats won or lost by the President's party in Congress based on the approval rating of the United States President.

However, this model could not be used to predict the outcome of the 2014 Midterm Elections for the Democratic Party because President Obama's approval rating is below 50%. That being said, if a President's approval rating is above 50%, this model is even more accurate than the originally calculated model.

**Now to find a model to predict seat gain when the President's approval rating is under 50%**

| Year | President | Party | President's Approval Rating Prior to Midterm (Percentage) | Seat Gain/Loss for President's party |
|---|---|---|---|---|
| 1978 | Carter | Democrat | 49 | -11 |
| 1994 | Clinton | Democrat | 46 | -53 |
| 2010 | Obama | Democrat | 45 | -63 |
| 1966 | Johnson | Democrat | 44 | -47 |
| 1982 | Reagan | Republican | 42 | -28 |
| 1950 | Truman | Democrat | 39 | -29 |
| 2006 | G. W. Bush | Republican | 38 | -30 |
| 1946 | Truman | Democrat | 33 | -55 |
| | | Mean | 42 | -39.5 |
| | | St. Deviation | 5.126959556 | 17.63114128 |

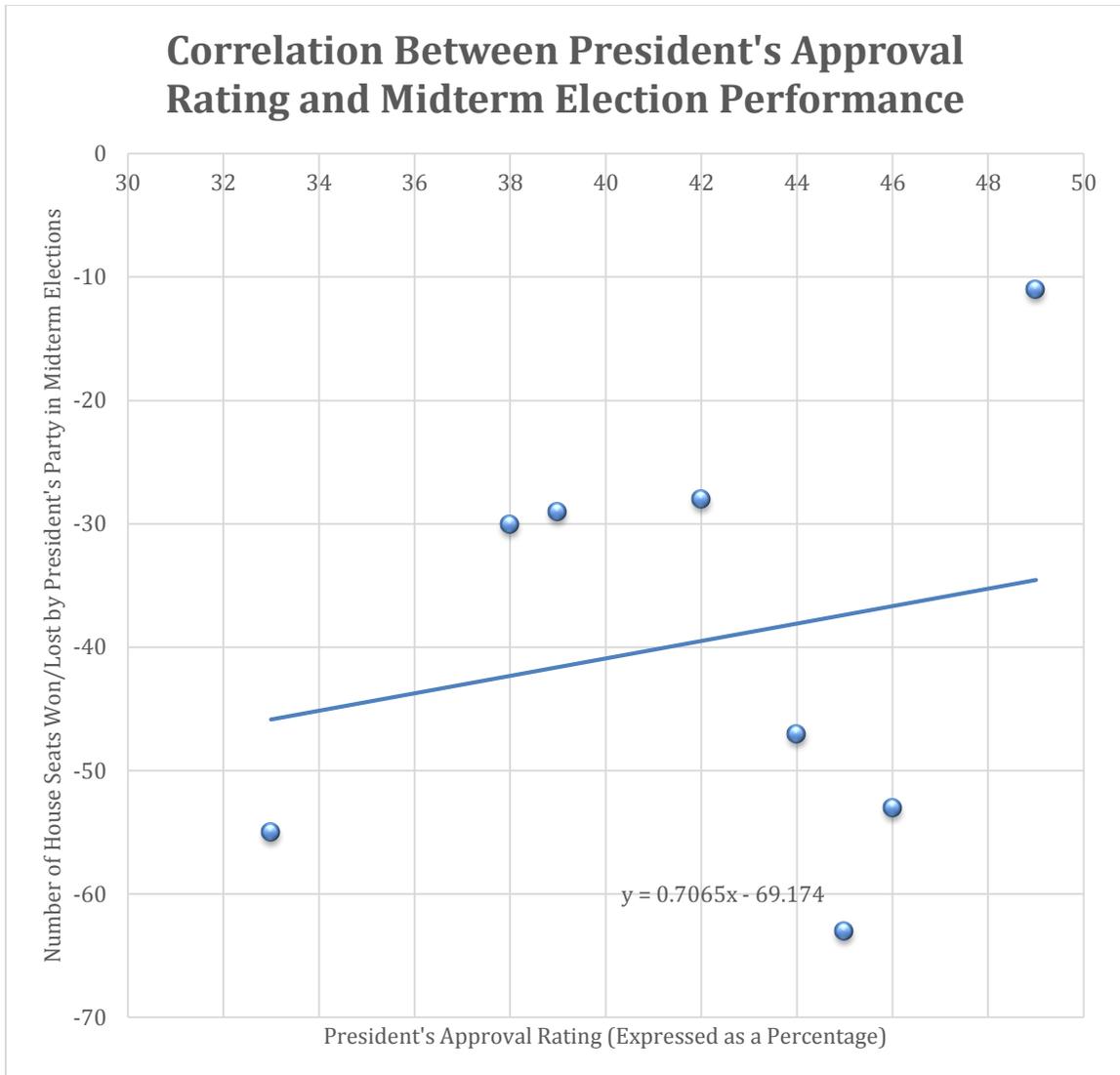

From these data, a few things are immediately clear.

1) The President's Party never gained seats during Midterms when his approval rating is under 50%

2) The President's approval rating is centered at about 42% with a standard deviation of about 5.12%. This means that the approval ratings are predominantly in the upper 30s to upper 40s range.

**Now to find a model for these data:**

| Linear Regression for Approval Ratings Less than 50% | | | | | |
|---|---|---|---|---|---|
| a | bx | $r^2$ | df | t | p |
| -69.1739 | 0.7065*x | 0.042269 | 6 | 0.512416 | 0.625485 |

The regression line for these data is y = -69.1739+0.7065x, with a correlation coefficient of about 0.042, showing a little to no correlation. This correlation coefficient clearly shows that this model is not useful, but a t test for the utility of the linear model will be done for confirmation. The hypotheses are once again as follows:

**$H_0$: The regression line is an improper fit ($\beta=0$)**
**$H_a$: $H_0$ is false**

The above linear regression analysis gives a t critical value of 0.512416 and a corresponding p value of 0.625485. Because the p value is so high, the null hypothesis, $H_0$, can be only be rejected about 37.5% confidence.

The line y = -69.1739+0.7065x is not a useful model for predicting the seats won or lost by the President's party in Congress during midterms. The original line, y = -107.423+1.594x, should be used to predict the seat gain by the President's party during Midterm Elections.

## CONCLUSION

From this analysis a statistically useful model, from which inferences about future midterm elections can be made, was found. The data regarding a United States President's approval rating and the number of Congressional seats won or lost by the President's political party during midterm elections was fitted to the line y = -107.423+1.594x. A test of utility proved with 99.8% certainty the usefulness of the above linear model. Said linear model was used to determine with 95% confidence that the Democratic Party during the 2014 Midterm Elections will lose between 48 and 27 seats based on President Obama's approval rating of 44% as of May 16, 2014. Further investigations showed an approximately 20% more useful predictor model for the midterm performance by the President's party when the President's approval rating is greater than 50%. There was no useful predictor model found exclusively for approval ratings less than 50%; the only conclusion that can be drawn for approval ratings under 50% is that the President's party will likely lose seats during midterm elections. During the review process, the 2014 Midterm Elections occurred and comparisons could be made between predictions and actual results. The result of the 2014 Midterm Elections, in which the Democrats lost 28 seats, validated this model's prediction.

This model can be powerful for political advisors and analysts, as it gives a predictor for how the President's approval rating influences how his party does in Midterm Elections. These findings also demonstrate a potential conundrum for a President's administration. A President will have a much easier time passing legislature if he has a majority of his party in Congress. But if a President cannot pass legislature, his approval rating can potentially go down. And if the President's approval rating goes down, his party will have a harder time winning seats in Congress. If the President's party loses a majority in Congress, he will have trouble passing legislature, which will bring down his approval rating, further potentially limiting his party's performance during midterm elections. Using this model, political advisors can recognize this conundrum and a President's administration can take steps to bolster the President's approval rating.

**Works Cited**

**15**"Avg. Midterm Seat Loss 36 for Presidents Below 50% Approval." *Avg. Midterm Seat Loss 36 for Presidents Below 50% Approval*. Gallup, 9 Aug. 2010. Web. 17 May 2014. <http://www.gallup.com/poll/141812/avg-midterm-seat-loss-presidents-below-approval.aspx>.

"Gallup Daily: Obama Job Approval." *Gallup Daily: Obama Job Approval*. Gallup, 16 May 2014. Web. 17 May 2014. <http://www.gallup.com/poll/113980/Gallup-Daily-Obama-Job-Approval.aspx>.